\documentclass[preprint,preprintnumbers,amsmath,amssymb,superscriptaddress]{revtex4-1}
\usepackage{fullpage,amsthm,amsmath,amsfonts,bm,times,color,bbm}
\usepackage{graphicx,subfigure}
\usepackage[english]{babel}
\usepackage{mathtools}

\def\eqa{\begin{eqnarray}}
\def\eea{\end{eqnarray}}
\newcommand{\eq}{\begin{equation}}
\newcommand{\ee}{\end{equation}}

\makeatother

\makeatletter

\begin{document}

\title{ 
Percolation Framework of the  Earth's Topography
}

\author{Jingfang Fan}
\email{jingfang@pik-potsdam.de}
\affiliation{Potsdam Institute for Climate Impact Research, 14412 Potsdam, Germany}
\author{Jun Meng}
\email{meng@pik-potsdam.de}
\affiliation{Potsdam Institute for Climate Impact Research, 14412 Potsdam, Germany}
\author{Abbas Ali Saberi}
\email{ab.saberi@ut.ac.ir}
\affiliation{Department of Physics, University of Tehran, Tehran 14395-547, Iran}
\affiliation{School of Particles and Accelerators, Institute for Research in Fundamental Sciences IPM, Tehran 14395-547, Iran}
 \affiliation{Institut f\"ur Theoretische Physik, Universitat zu K\"oln, Z\"ulpicher Strasse 77, 50937 K\"oln, 
 	Germany}

\begin{abstract}

Self-similarity and long-range correlations are the remarkable features of the Earth’s surface topography. Here
we develop an approach based on percolation theory to study the geometrical features of Earth. Our analysis is
based on high-resolution, 1 arc min, ETOPO1 global relief records.We find some evidence for abrupt transitions
that occurred during the evolution of the Earth’s relief network, indicative of a continental/cluster aggregation.
We apply finite-size-scaling analysis based on a coarse-graining procedure to show that the observed transition
is most likely discontinuous. Furthermore, we study the percolation on two-dimensional fractional Brownian
motion surfaces with Hurst exponent $H$ as a model of long-range correlated topography, which suggests that the
long-range correlations may play a key role in the observed discontinuity on Earth. Our framework presented
here provides a theoretical model to better understand the geometrical phase transition on Earth, and it also
identifies the critical nodes that will be more exposed to global climate change in the Earth’s relief network.

\end{abstract}
\date{\today}

\maketitle
\section{INTRODUCTION}

The topography or bathymetry of Earth shows complex multifractal structures and scaling properties~\cite{mandelbrot_stochastic_1975,gagnon_multifractal_2006,sapoval_self-stabilized_2004,maritan_universality_1996}, which can be considered as a consequence of plate tectonic processes. Despite most of the major surface topographic features of Earth can be explained by the plate tectonic theory \cite{gill2012orogenic}, little is known about how to identify and detect critical geographical positions and geology regions on the Earth's surface. 
The surface topography of Earth plays a remarkable role in the dynamical evolution of oceans, especially with reference to the global climate changes and sea level rising.

Network science has demonstrated its potential as a useful tool in the study of real world systems ranging from physics and biology to the social systems~\cite{albert_statistical_2002,cohen2010complex,brockmann_hidden_2013,newman2010networks,romualdo_pastor-satorras_epidemic_2015}. It has also been successfully applied in climate sciences to construct ``climate networks'', in which the geographical locations are regarded as nodes, the similarity between the records of different nodes represents the links~\cite{tsonis_topology_2008,yamasaki_climate_2008,donges_complex_2009,donges_backbone_2009,barreiro_inferring_2011,martin_interpretation_2013,fan2017network}. Climate networks were used to forecast climate extreme events, such as El Ni\~{n}o and heavy rainfall \cite{ludescher_very_2014,boers_prediction_2014,meng_percolation_2017,meng_forecasting_2018}. Detecting and identifying vital and critical nodes in networks plays a significant role in unveiling fundamental organization principles of complex systems~\cite{runge_identifying_2015,kitsak_identification_2010,liu_controllability_2011,lu_vital_2016}.

Percolation theory is an effective tool for understanding the resilience of connected clusters to node breakdowns through topological and structural properties \cite{isichenko_percolation_1992,bunde2012fractals,cohen_resilience_2000,aharony2003introduction}. The essence of the analysis is the identification of a system's different clusters and the connectivity between them. It has been applied to many natural and human-made systems \cite{taubert_global_2018,ali_saberi_percolation_2013,li_percolation_2015,morone_influence_2015,meng_percolation_2017,saberi2015recent}. Here, we combine network and percolation theory to identify and detect critical nodes in Earth's relief network, we find abrupt percolation transitions occurred during the evolution of the network. We then discuss the nature of geometric phase transitions by using the finite size scaling theory. Furthermore, we detect the critical nodes in  Earth's relief network.


\section{Data and Methods}
\subsection{Data}
In this study, we use the topographic data of ETOPO1 Global Relief Model. It is used to calculate the Volumes of the World's Oceans and to derive a Hypsographic Curve of Earth's Surface, built from global and regional data sets \cite{amante1noaa}. The resolution is 1 arc-minute, i.e., $N = 10800 \times 21600$ grid points. The present mean sea level (zero height) is assumed as a vertical datum of the height relief $h(\phi_i,\theta_i)$, where $\phi_i$, $\theta_i$  are the corresponding latitude and longitude of grid point $i$.
The data can be downloaded from \url{https://www.ngdc.noaa.gov/mgg/global/global.html}.

\subsection{Methods}

In percolation on a lattice system, each lattice site (or bond) is occupied with probability $p$. A set of occupied sites in which every site is connected to its nearest neighbors forms a distinct cluster. We first rank all the grid points in the ETOPO1 Global Relief Model according to their height $h(\phi_i,\theta_i)$, from the largest to the smallest value. A number is then assigned to each site according to the position of its height in the rank. The percolation model is defined as follows: starting from an unoccupied lattice, the sites are occupied one by one according to their ranking, i.e., we first choose the site with the highest height, then the second and so on. At each step, the fraction of occupied sites $p$ increases by the inverse of the total number of sites $N$ in the Earth's relief landscape. By this procedure, a configuration of occupied sites is continuously obtained at every $p$. 
Our method here is different from the one presented in Ref. \cite{ali_saberi_percolation_2013} in which the control parameter is the height level with a rather less resolution. The advantage of our approach here is that it enables  us to exactly identify the critical nodes on the Earth's topography.

We then identify the clusters based on classical graph theory: a cluster is a subset of network nodes such that there exists at least one path from each node in the subset to another \cite{cohen2010complex,newman2010networks}. To detect the clusters in evolving lattice system, one can use either the Hoshen--Kopelman algorithm \cite{hoshen_percolation_1976} or the efficient Newman--Ziff algorithm \cite{newman_efficient_2000}. We apply periodic boundary conditions along the zonal direction, and free boundary conditions along the meridional direction. Each node has indeed four nearest neighbors. We denote $G_m$ as a series of sub-networks. The order parameter of percolation in the network/lattice systems is defined as the relative size of the largest cluster \cite{cohen2010complex}. Due to the Earth's spherical shape, the largest cluster in the spatial relief network is defined as \cite{fan2018climate}, 
\begin{equation}
s = \frac{\max \left[\sum\limits_{i\in G_1} \cos(\phi_i),\cdots, \sum\limits_{i\in G_m} \cos(\phi_i),\cdots,\right]}{\sum\limits_{i=1}^{N} \cos(\phi_i)}. 
\label{eq1}
\end{equation} 
The system on a regular lattice is considered as percolating if there is a path from one side of the lattice to the opposite side, passing only through the occupied links/nodes. When such a path exists, the component or cluster of sites that spans the network from side to side is called the spanning cluster \cite{cohen2010complex}. However, in many systems such as on networks, no notion of ``side'' exists. For example in our case here for Earth, in zonal direction one cannot define the sides, however, when one looks at the behavior of the largest component containing $\mathcal{O}(N)$ nodes/links, there will be a divergent correlation length and mean-island size at the largest gap in the order parameter \cite{ali_saberi_percolation_2013}. 



\section{RESULTS}

\subsection{Earth's topography}
\subsubsection{Landmass percolation clusters}

\begin{figure}
\begin{centering}
\includegraphics[width=0.9\linewidth]{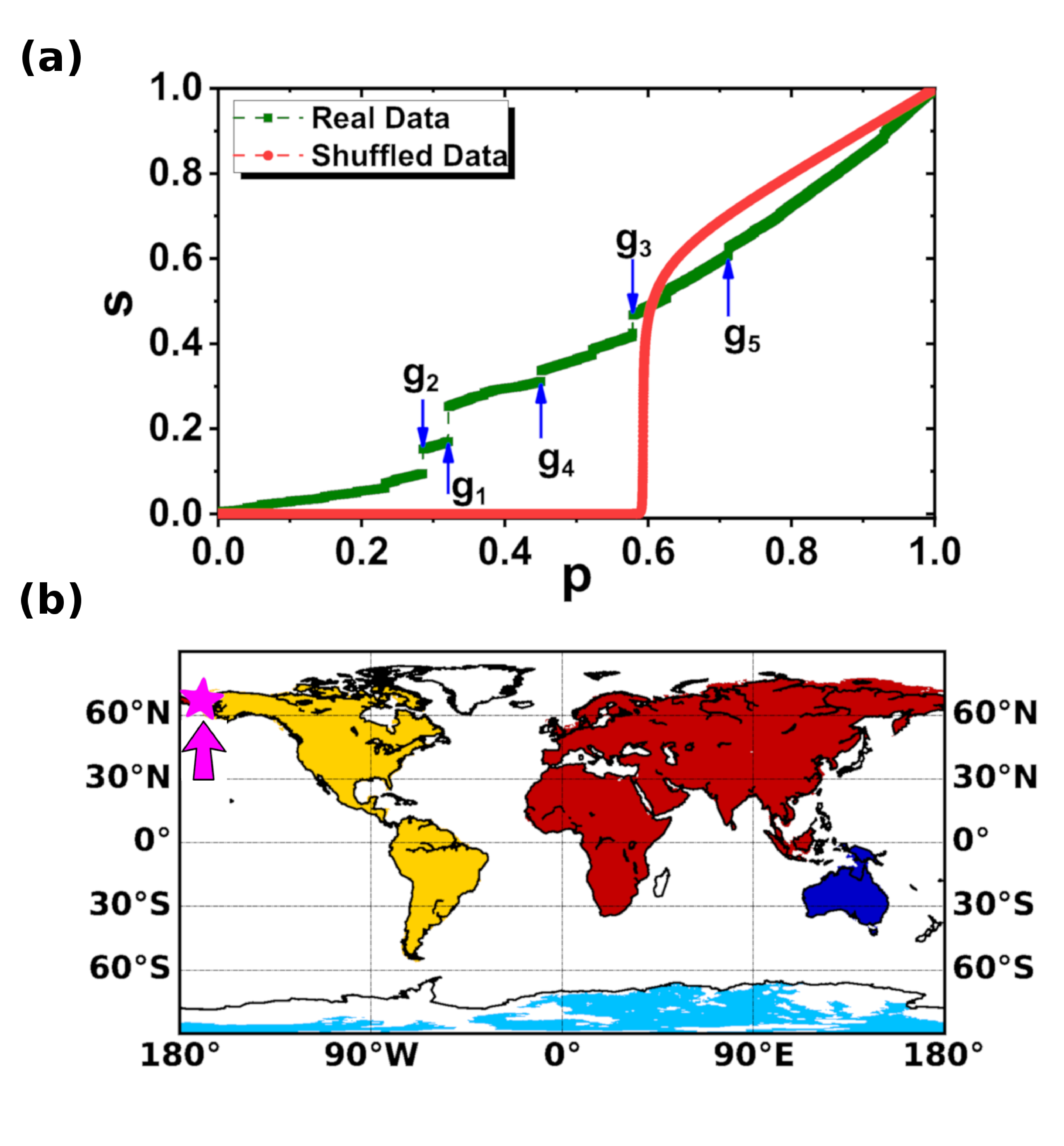}
\caption{\label{Fig_1} (a) The largest landmass cluster relative size $s$ versus the fraction total landmass, $p$, for real (square green) and shuffled (circle red) Earth's relief records. $g_1$--$g_5$ indicate the largest five gaps, defined in Eq.~\ref{eq2}. (b) Snapshots of the landmass clusters  of the Earth's surface topography network just before the percolation threshold (the largest jump  at $p\approx0.321$). Different colors represent different clusters; the grid resolution is 1 arc-minute; the star indicates the critical node. Only the clusters with relative size larger than $0.01$ are shown.}
\end{centering}
\end{figure}

The topography of Earth is complex and its morphology is a result of diverse processes such as tectonic activity and erosion. 
Similar to our prior work \cite{ali_saberi_percolation_2013}, we study the Earth's topography by means of the percolation theory. Our evolving spatial relief network starts globally with $N$ isolate nodes, the nodes are occupied one by one according to their height $h$ [see more details in Section Methods]. 
We then analyze the dynamical evolution of the largest cluster $s$ as a function of the fraction of occupied nodes $p$. As shown in Fig.~\ref{Fig_1}(a), we find that Earth's relief network undergoes several abrupt and statistically significant phase transitions, i.e., exhibiting a significant discontinuity in the order parameter $s$. The size of the $i$-th gap $g_i$ at each fraction $p$ is defined as follows:
\begin{equation}
g_i(p) \equiv s(p)-s(p-1/N).
\label{eq2}
\end{equation}
Therefore, $g_1$ indicates the largest gap, $g_2$ the second largest gap, and so forth.
The larger the gap $g_i$, the larger the two clusters before merger. Therefore, the largest gap $g_1$ in the order parameter is a natural candidate for a possible percolation transition and formation of a giant component of $\mathcal{O}(N)$ nodes \cite{fan2018climate}.

Fig.~\ref{Fig_1}(b) shows the network landmass clusters structure in
the Earth's surface map at the percolation threshold (just before the largest gap that is indicated by the blue arrow with $g_1$). We find that the network, just before this jump, is characterized by four major communities ---only clusters with size larger than 0.01 are shown--- the largest one is the Afro-Eurasia  continental landmass (indicated by red color); the second largest cluster is the Americas (indicated by yellow color); the third one is located in the Antarctica and the fourth is the Oceania. A critical node $(64.458333~\hbox{$^\circ$}N,171.141667~\hbox{$^\circ$}W)$ connects the largest and second largest cluster at the percolation threshold $p_c \approx 0.321$, with altitude level $h=-43$ m, under the current sea level [see Fig.~S1].

\subsubsection{Oceanic percolation clusters}

\begin{figure}
\begin{centering}
\includegraphics[width=1.0\linewidth]{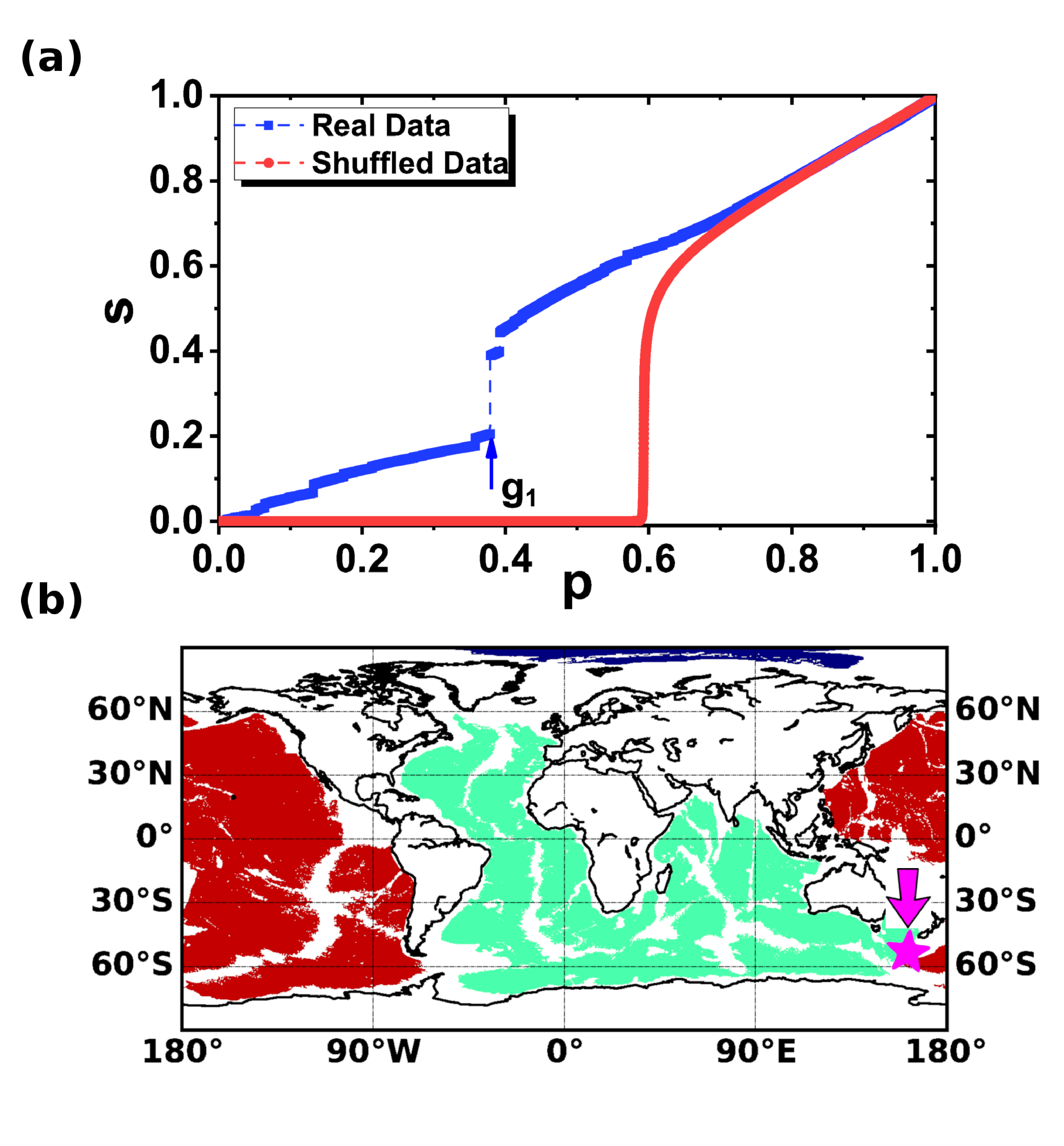}
\caption{\label{Fig_5} Similar to Fig.~\ref{Fig_1} but for the oceanic clusters. (a) The largest oceanic cluster relative size $s$ versus $p$. (b) Snapshots of the oceanic clusters structure of the Earth's relief network just before the percolation threshold (the largest jump  at $p\approx 0.379$). The star indicates the critical node.}
\end{centering}
\end{figure}

The same analysis can be applied to characterize the oceanic clusters, i.e., the nodes are added one by one according to their hight level in increasing order. As shown in Fig.~\ref{Fig_5}(a), it also gives rise to a discontinuous jump in the oceanic order parameter at the percolation threshold $p_c \approx 0.379$, with altitude level, $h=-3621$ m. This is very close to the result, $h=-3640$ m, reported in Ref. \cite{ali_saberi_percolation_2013}. The network oceanic clusters, just before the jump are shown in 
Fig.~\ref{Fig_5}(b). We find that the critical node, $(59.908333~\hbox{$^\circ$}S, 161.308333~\hbox{$^\circ$}E)$, connects the Atlantic+Indian Ocean Plate (colored in red) to the Pacific Plate (colored in yellow). 
It is worth noting that the role of the largest cluster on the landmass structure is different from that on the oceanic one, e.g., the percolation threshold as well as the corresponding critical nodes are different. These differences unveil the complex and different structure of the Earth's topography (continents) and bathymetry (oceans). This dichotomy is manifest in the well-known bimodal distribution of the Earth's topography~\cite{wegener1966origin}.

\subsubsection{Order of the percolation transition}

In order to demonstrate that the observation of the jumps in the order parameter are not accidental, we analyze randomized topography obtained 
from the shuffling of the original data. This procedure removes the long-range correlations in the height profile while keeps the height distribution intact. We have considered 100 such randomizations and determined the averaged giant cluster $s$, as shown in Fig.~\ref{Fig_1}(a). 
We also checked that the behavior for the shuffled data is independent of the realizations and the same result obtains for a single realization as well (see Fig. S2 \cite{SI}). As expected, the shuffled samples all correspond to the classical uncorrelated site percolation class \cite{newman_efficient_2000} with a continuous phase transition at $p\sim 0.59$.  


\begin{figure}
\begin{centering}
\includegraphics[width=1.0\linewidth]{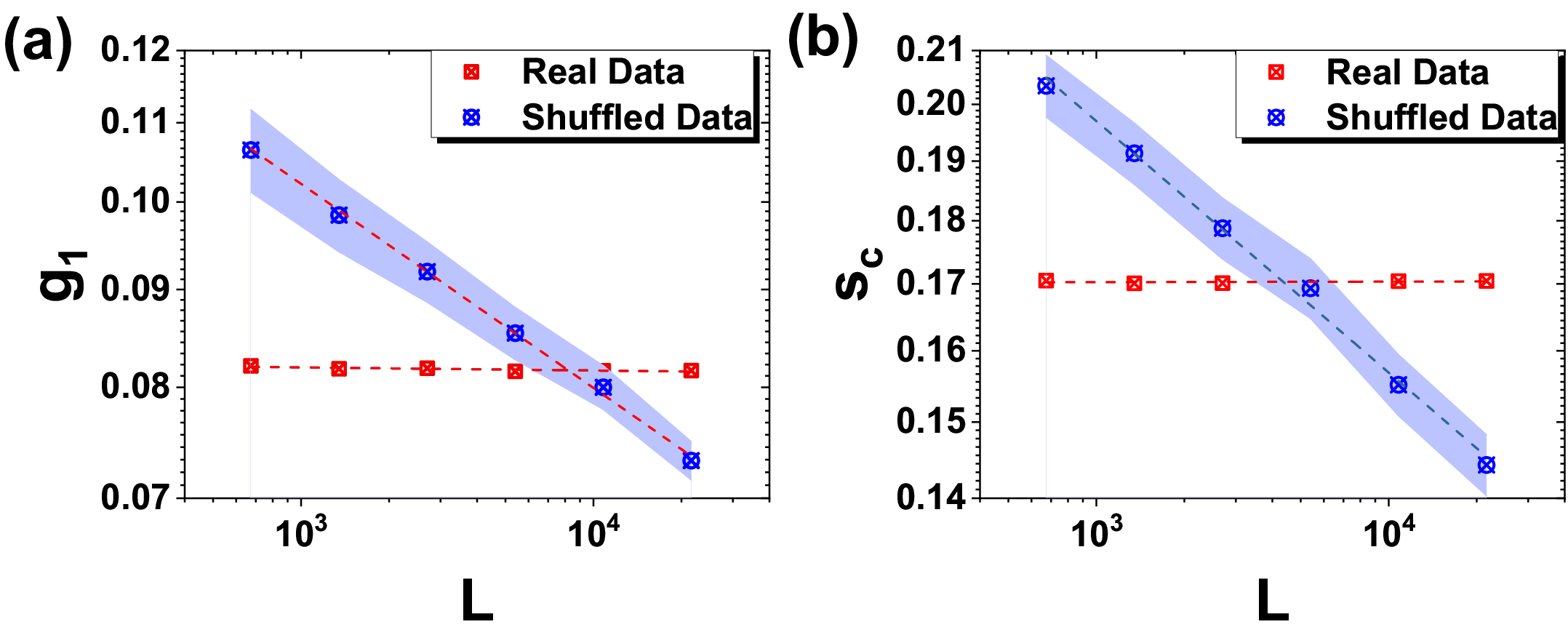}
\caption{\label{Fig_4} Finite size effects of the percolation in Earth's relief network. (a) Log-log plot of the largest gap $g_1$ versus the network system size $L$ for original real data (square red) and shuffled data (circle blue). (b) Log-log plot of the largest landmass cluster relative size $s_c$ at the percolation threshold, versus $L$ for real data (square red) and shuffled data (circle blue). For the real data, the slope seems to approach zero, suggesting a discontinuous phase transition; for the shuffled data, the slope approaches -0.10, which suggests a continuous phase transition with a known critical exponent $\beta/\nu$ = 5/48 and $d - d_f$ = 5/48. The dashed lines are the best fit-lines with R-Square $>$ 0.98. The shaded regions correspond to error bars, which are calculated by the standard deviation.}
\end{centering}
\end{figure}

It has been pointed out that a random network or lattice system always undergoes a continuous percolation phase transition and shows standard scaling features during a random process ~\cite{bollobas2001random}. The question whether percolation transitions could be discontinuous has attracted much attention recently in the context of interdependent networks~\cite{buldyrev_catastrophic_2010,hu_percolation_2011,gao_networks_2012} and the so-called explosive percolation models ~\cite{achlioptas_explosive_2009,riordan_explosive_2011,fan_continuous_2012,dsouza_anomalous_2015}. Interestingly, the dynamic evolution on our Earth's relief network indicates the possibility of discontinuous phase transition, as shown in Fig~\ref{Fig_1}(a). To further test the order of the percolation phase transition, we study the finite size effects of our network by altering the resolution of nodes. We then calculate $g_1 (L)$, the largest gap in $s$ as a function of network system size $L$ and see how it behaves when extrapolated to the infinite system size. $L$ is defined as the number of nodes in zonal direction. If $g_1 (L)$ approaches zero as $L \to \infty$, the corresponding giant cluster is assumed to undergo a continuous percolation; otherwise, the corresponding percolation is assumed to be discontinuous \cite{nagler_impact_2011}. The results are shown in Fig.~\ref{Fig_4}(a). It suggests a discontinuous percolation since $g_1 (L)$ tends to be a non-zero constant. For comparison, we also show the continuous results for shuffled data, with a known critical exponents $\beta/\nu = 5/48 \approx 0.104$ \cite{aharony2003introduction}. Where $\beta$ is the critical exponent of the order parameter $s \sim |p - p_c|^{\beta}$, and $\nu$ describes the divergence of the correlation length $\xi \sim |p - p_c|^{-\nu}$. 
In addition, we examine the scaling relation with size for the order parameter at the percolation threshold (just before the largest jump $g_1$),
$s_c \sim L^{d_{f}-d}$, for both real and shuffled data (see Fig.~\ref{Fig_4}(b)). We find that $d_{f} -d = 0$, an indicative of discontinuous percolation for our real network; however, for the shuffled case from simulations, $d_{f} -d \approx -0.104$, which agrees well with the known exponent value for standard percolation in two dimensions, i.e.,  $d_f = 91/48$~\cite{aharony2003introduction,bunde2012fractals}. The dashed lines shown in Fig.~\ref{Fig_4} are the best fit-lines for the data with R-Square $>0.98$.

Fig.~S3 presents the order parameter, $s$, as a function of occupied probability $p$ with different system size $L$. We find that there are no significant finite-size effects for our system since the four curves with $L = 21600, 10800, 5400$ and $2700$ are nearly overlapping. The network landmass clusters structure at the $5$ percolation thresholds (corresponding to the largest $5$ jumps) with $L = 2700$ is shown in Fig.~S4. We find that they have the similar manners compared to the cases with $L = 21600$ [Fig.~\ref{Fig_1}(b)], which indicates the self-similar fractal patterns  of Earth's surface topography \cite{mandelbrot_stochastic_1975,mandelbrot_how_1967}.  

\subsection{Origin of the discontinuity}

\begin{figure}
\begin{centering}
\includegraphics[width=1.0\linewidth]{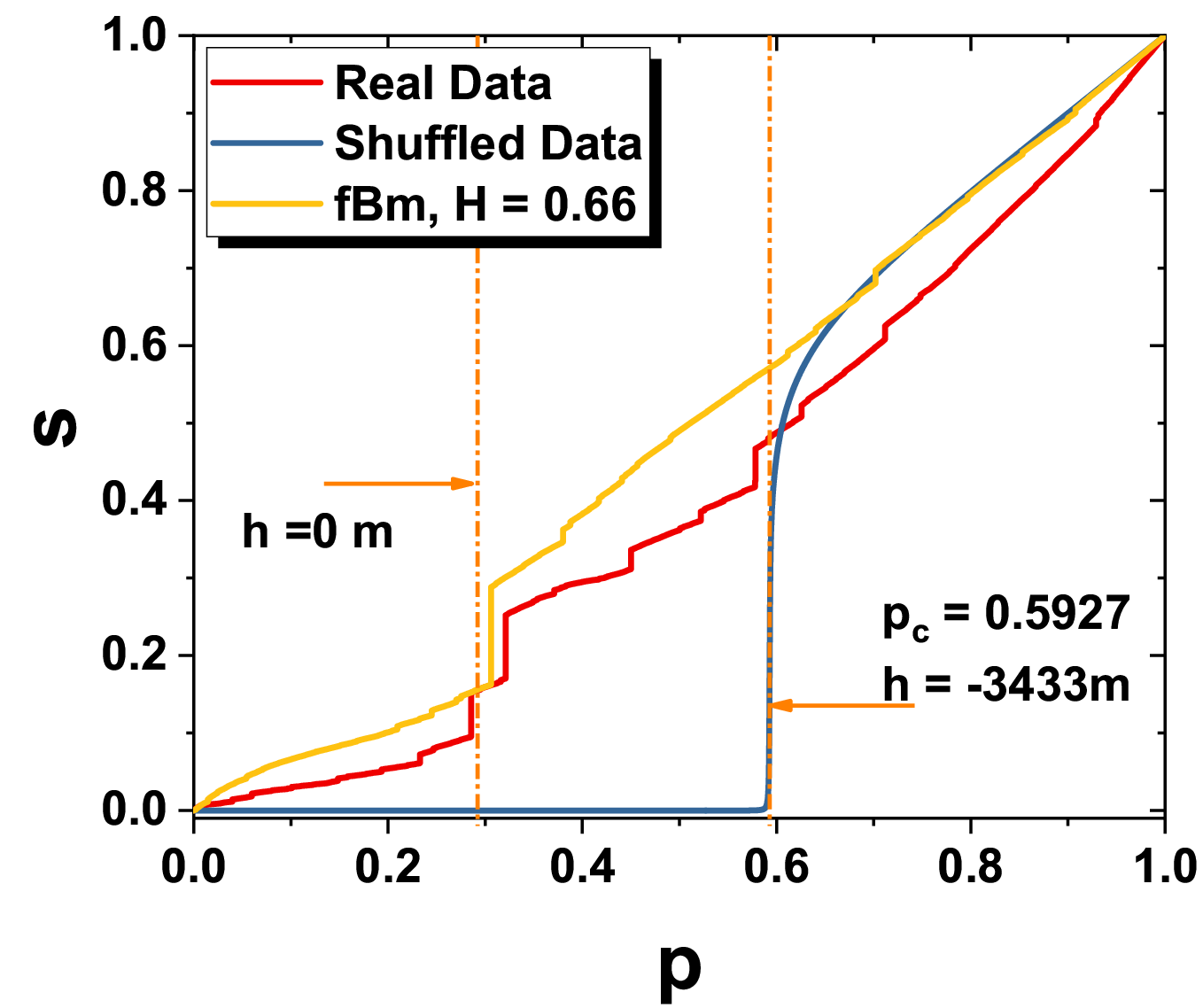}
\caption{\label{Fig_51} The largest landmass cluster relative size $s$ versus the fraction number of nodes/sites, $p$, for real, shuffled and 2d fractional Brownian motion (fBm) surfaces with different Hurst exponents. The two dashed lines indicate the sea level (h=0 m) and the well known site percolation threshold $p_c = 0.5927$, respectively.}
\end{centering}
\end{figure}

\begin{figure}
\begin{centering}
\includegraphics[width=1.0\linewidth]{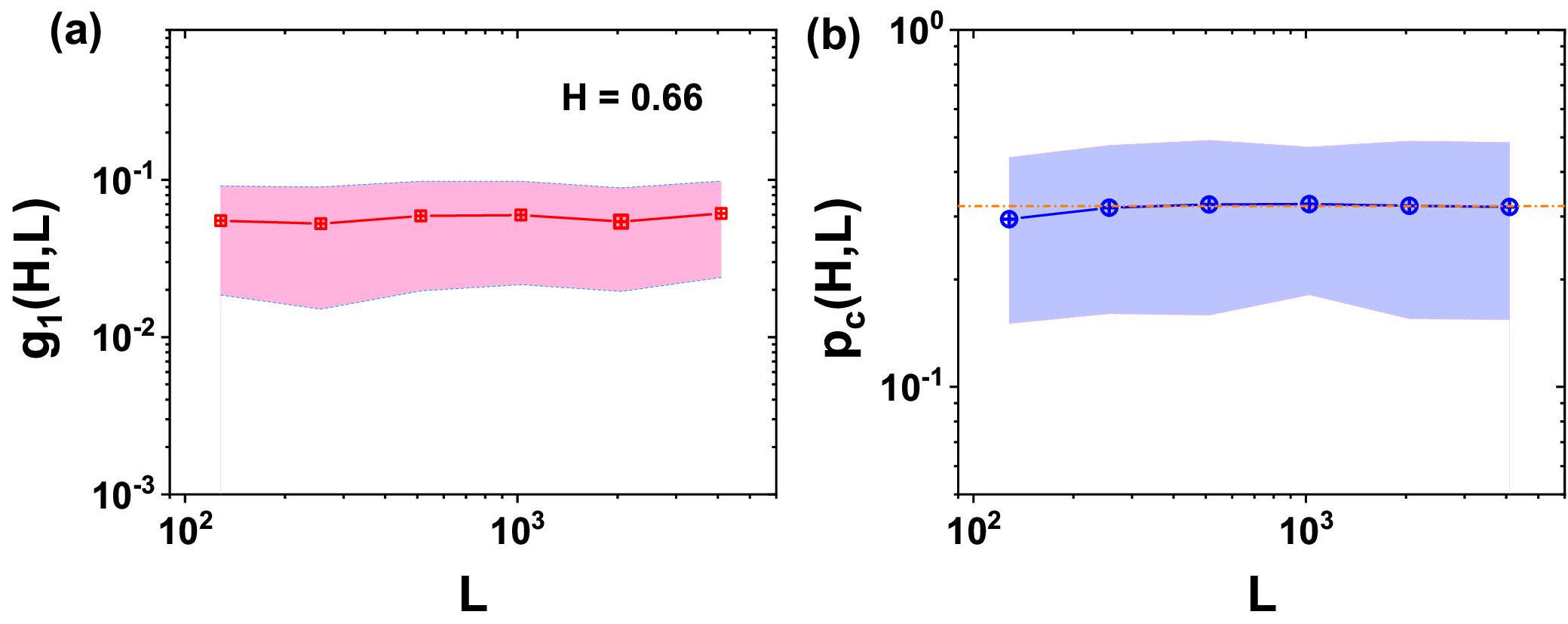}
\caption{\label{Fig_52} The percolation on 2d fractional Brownian motion surfaces. (a) The average of the largest jump $g_{1}(H,L)$ as a function of system size $L$; (b) the corresponding percolation threshold $p_c (H,L)$ as a function of $L$. The dashed line in (b) stands for the percolation threshold $p_c = 0.321$ for real data. Here, $H=0.66$ is the Hurst exponent governing the correlations over the continental topography, following ~\cite{gagnon_multifractal_2006}. The shaded regions correspond to error bars, estimated by the standard deviation.}
\end{centering}
\end{figure}

To better understand the Earth's topography and the origin of the discontinuity, we study the percolation on 2d fractional Brownian motion (fBm) surfaces with Hurst exponent $H$ \cite{du_percolation_1996,s1992fractal}. The parameter $H$ is usually between 0 and
1, where 0 is very noisy, and 1 is smoother. The Earth's rough surfaces can be modeled by fractional Brownian motion \cite{family1991dynamics}, and the estimation over the continental topography is $H = 0.66$ \cite{gagnon_multifractal_2006}. We present the percolation analysis results on fBm surfaces with $H = 0.66$ in Fig.~\ref{Fig_51}, and the corresponding 2d fBm surface is shown in Fig.~S5. Similar to the real network, we find that $s$ also exhibits abrupt transitions around $p$ whose positions are sample dependent.
To further test the order of the percolation phase transitions on fBm surfaces, we then use the finite size scaling theory.  The largest gap $g_1(H,L)$ (average) as a function of system size $L$ are shown in Fig.~\ref{Fig_52}(a). We find that percolation on fBm surfaces with $H=0.66$ is discontinuous, since $g_1(H,L)$ tends to be a non-zero constant for the very large extrapolated system size. We also find that the location of the threshold is size-independent---Fig.~\ref{Fig_52}(b). 
%
This indicates that the percolation method can be used as an efficient tool to study the Earth's topography.
It has recently been shown that the percolation with Hurst exponents $H\in [-1,0]$ is continuous~ \cite{castro_schramm-loewner_2018,schrenk_percolation_2013}; and our results suggest that the percolation on fBm with Hurst exponents $H>0$ is discontinuous.
Figs.~S6--S8 show our three specific examples of fBm surfaces and their corresponding percolation process  with $H = 0.1, 0.5$ and $0.9$, respectively. 
The origin of the discontinuity might be explained by the plate tectonic theory (or the  long-range correlated
 behavior of the Earth's  topography).


%
%

\subsection{Critical Nodes in Earth's relief network}

\begin{figure}
\begin{centering}
\includegraphics[width=1.0\linewidth]{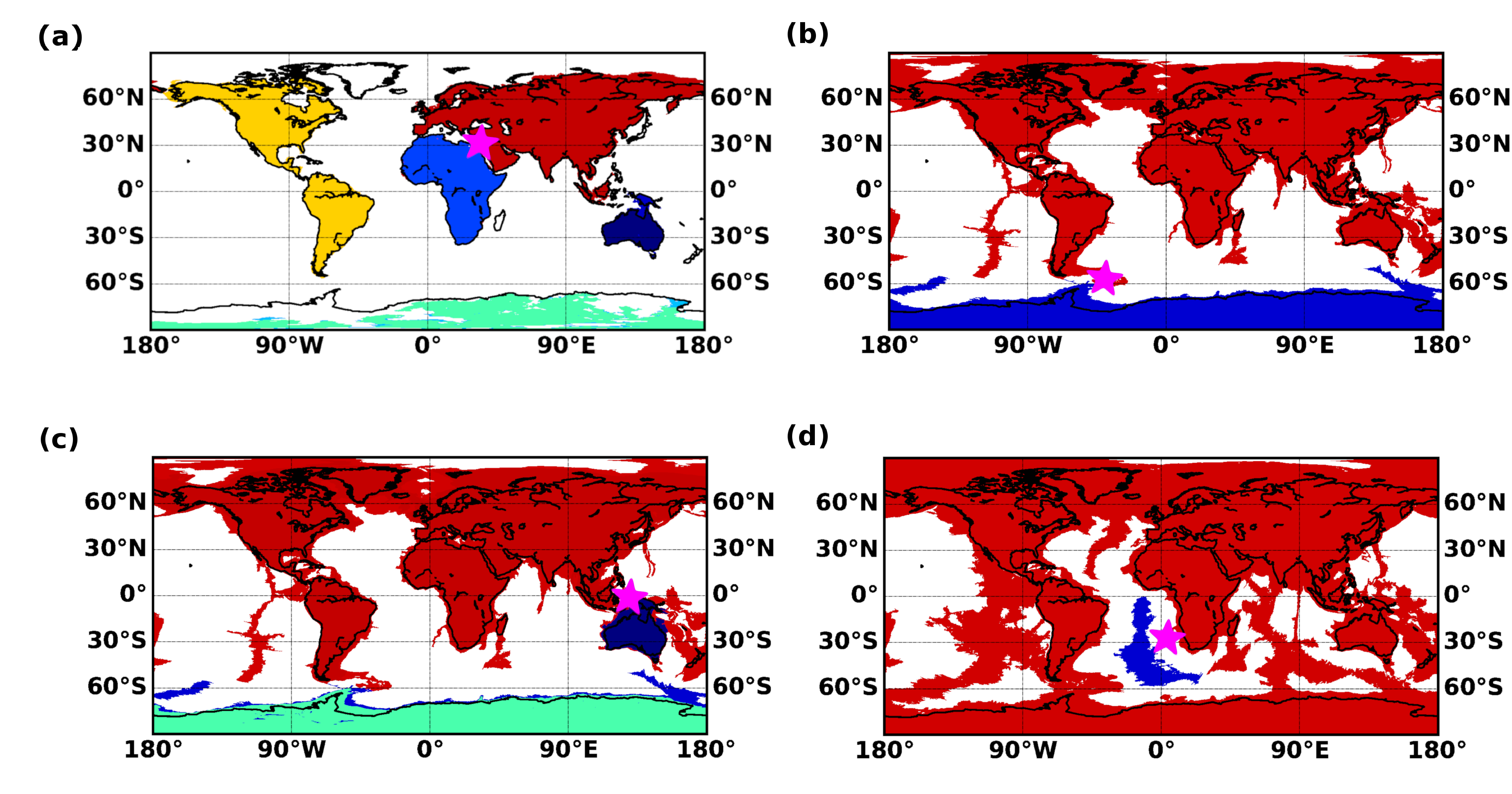}
\caption{\label{Fig_3} Snapshots of the landmass clusters structure of the Earth relief network just before the percolation threshold at (a) the second largest jump with $p\approx 0.285$; (b) the third largest jump with $p\approx 0.578$; (c) the fourth largest jump with $p\approx 0.450$; (d) the fifth largest jump with $p\approx 0.712$. The red color represents the largest landmass cluster, the stars indicate the critical nodes.}
\end{centering}
\end{figure}

clusters experience some abrupt jumps which can be indicative of a percolation transition that happens at some critical and vital nodes which are responsible for the connectedness of two major parts. In this study, for simplicity and without loss of generalization, we chose the largest five gaps during the evolution of our spatial network [see Fig.~\ref{Fig_1}(a)] to identify the critical nodes. These ``critical nodes'' are crucial for future global climate regime changes \cite{cathcart2010bering}. In the previous chapters, we have detected one critical node corresponding to the largest gap [see Fig.~\ref{Fig_1} (b) and Fig.~\ref{Fig_5} (b)]. In the following, we perform the same analysis for the other four smaller jumps, $g_2$, $g_3$, $g_4$ and $g_5$ [as shown in Fig~\ref{Fig_1}(a)]. The corresponding network landmass clusters structure are presented in Fig.~\ref{Fig_3}. The critical nodes 
are: $(30.541667~\hbox{$^\circ$}N, 32.125~\hbox{$^\circ$}E)$  with altitude level $h=12$ m; $(60.575~\hbox{$^\circ$}S, 39.258333~\hbox{$^\circ$}W)$ with altitude level $h=-3325$ m;
$(8.541667~\hbox{$^\circ$}S, 129.208333~\hbox{$^\circ$}E)$ with altitude level $h=-1292$ m and $(32.425~\hbox{$^\circ$}S, 1.641667~\hbox{$^\circ$}W)$ with altitude level $h=-4116$ m, respectively.

\section{SUMMARY}

The Earth's topography has been considered as some kinds of correlated fields as described by Isichenko in \cite{isichenko_percolation_1992}.  However, one should notice that there exists a number of differences that makes the Earth's topography different with characteristic properties. For example, unlike an fBm surface, Earth has a bimodal distribution of heights with different Hurst exponents giving rise to distinct scaling properties over oceans, continents, and continental margins \cite{gagnon_multifractal_2006}. In the present study, we have applied $H=0.66$ known for the continental topography.  Moreover, our numerical results for the site percolation of an fBm landscape highly agree with the results reported in \cite{du_percolation_1996}, i.e., the percolation threshold $p_c$ does not change with the system size
(see Fig.~\ref{Fig_52}(b) and Table I in \cite{du_percolation_1996}). 
The results shown in Fig.~\ref{Fig_52}(a) suggest that the percolation on fBm surfaces is discontinuous.
%

In summary, we have developed a framework to study the geometrical topography of Earth. Using  percolation analysis, we have studied the dynamical evolution of the giant landmass (oceanic) cluster, $s$, and found that the Earth's relief network exhibits abrupt transitions. We have used the discontinuous jumps of $s$ to detect the percolation threshold and identify the critical nodes over the globe. In addition, our finite-size analysis suggests that the observed jumps in the order parameter are most likely reminiscence of a discontinuous phase transition. To better understand the Earth's topography, we have analyzed the percolation properties on profiles generated by fBm surfaces. We have found that the presence of the long-range correlations among the height profile plays a key role in the characteristic  statistical behavior of the Earth's topography. 
The study of the Earth's network system may enrich our understanding of the statistical topography of Earth and can potentially be used as a template to study other complex systems.

\section*{Acknowledgements}
We acknowledge S. Havlin, Y. Ashkenazy and Richard Cathcart for their helpful
suggestions. J.F thanks the ``East Africa Peru India Climate Capacities --- EPICC'' project, which is part of the International Climate Initiative (IKI). The Federal Ministry for the Environment, Nature Conservation and Nuclear Safety (BMU) supports this initiative on the basis of a decision adopted by the German Bundestag. The Potsdam Institute for Climate Impact Research (PIK) is leading the execution of the project together with its project partners The Energy and Resources Institute (TERI) and the Deutscher Wetterdienst (DWD). A.A.S. would like to acknowledge
support from the Alexander von Humboldt Foundation
and partial financial support from the research council
of the University of Tehran.

\bibliographystyle{apsrev4-1}
\bibliography{MyLibrary}

\begin{thebibliography}{57}%
\makeatletter
\providecommand \@ifxundefined [1]{%
 \@ifx{#1\undefined}
}%
\providecommand \@ifnum [1]{%
 \ifnum #1\expandafter \@firstoftwo
 \else \expandafter \@secondoftwo
 \fi
}%
\providecommand \@ifx [1]{%
 \ifx #1\expandafter \@firstoftwo
 \else \expandafter \@secondoftwo
 \fi
}%
\providecommand \natexlab [1]{#1}%
\providecommand \enquote  [1]{``#1''}%
\providecommand \bibnamefont  [1]{#1}%
\providecommand \bibfnamefont [1]{#1}%
\providecommand \citenamefont [1]{#1}%
\providecommand \href@noop [0]{\@secondoftwo}%
\providecommand \href [0]{\begingroup \@sanitize@url \@href}%
\providecommand \@href[1]{\@@startlink{#1}\@@href}%
\providecommand \@@href[1]{\endgroup#1\@@endlink}%
\providecommand \@sanitize@url [0]{\catcode `\\12\catcode `\$12\catcode
  `\&12\catcode `\#12\catcode `\^12\catcode `\_12\catcode `\%12\relax}%
\providecommand \@@startlink[1]{}%
\providecommand \@@endlink[0]{}%
\providecommand \url  [0]{\begingroup\@sanitize@url \@url }%
\providecommand \@url [1]{\endgroup\@href {#1}{\urlprefix }}%
\providecommand \urlprefix  [0]{URL }%
\providecommand \Eprint [0]{\href }%
\providecommand \doibase [0]{http://dx.doi.org/}%
\providecommand \selectlanguage [0]{\@gobble}%
\providecommand \bibinfo  [0]{\@secondoftwo}%
\providecommand \bibfield  [0]{\@secondoftwo}%
\providecommand \translation [1]{[#1]}%
\providecommand \BibitemOpen [0]{}%
\providecommand \bibitemStop [0]{}%
\providecommand \bibitemNoStop [0]{.\EOS\space}%
\providecommand \EOS [0]{\spacefactor3000\relax}%
\providecommand \BibitemShut  [1]{\csname bibitem#1\endcsname}%
\let\auto@bib@innerbib\@empty
\bibitem [{\citenamefont {Mandelbrot}(1975)}]{mandelbrot_stochastic_1975}%
  \BibitemOpen
  \bibfield  {author} {\bibinfo {author} {\bibfnamefont {B.~B.}\ \bibnamefont
  {Mandelbrot}},\ }\href {\doibase 10.1073/pnas.72.10.3825} {\bibfield
  {journal} {\bibinfo  {journal} {Proceedings of the National Academy of
  Sciences}\ }\textbf {\bibinfo {volume} {72}},\ \bibinfo {pages} {3825}
  (\bibinfo {year} {1975})}\BibitemShut {NoStop}%
\bibitem [{\citenamefont {Gagnon}\ \emph {et~al.}(2006)\citenamefont {Gagnon},
  \citenamefont {Lovejoy},\ and\ \citenamefont
  {Schertzer}}]{gagnon_multifractal_2006}%
  \BibitemOpen
  \bibfield  {author} {\bibinfo {author} {\bibfnamefont {J.-S.}\ \bibnamefont
  {Gagnon}}, \bibinfo {author} {\bibfnamefont {S.}~\bibnamefont {Lovejoy}}, \
  and\ \bibinfo {author} {\bibfnamefont {D.}~\bibnamefont {Schertzer}},\ }\href
  {\doibase 10.5194/npg-13-541-2006} {\bibfield  {journal} {\bibinfo  {journal}
  {Nonlin. Processes Geophys.}\ }\textbf {\bibinfo {volume} {13}},\ \bibinfo
  {pages} {541} (\bibinfo {year} {2006})}\BibitemShut {NoStop}%
\bibitem [{\citenamefont {Sapoval}\ \emph {et~al.}(2004)\citenamefont
  {Sapoval}, \citenamefont {Baldassarri},\ and\ \citenamefont
  {Gabrielli}}]{sapoval_self-stabilized_2004}%
  \BibitemOpen
  \bibfield  {author} {\bibinfo {author} {\bibfnamefont {B.}~\bibnamefont
  {Sapoval}}, \bibinfo {author} {\bibfnamefont {A.}~\bibnamefont
  {Baldassarri}}, \ and\ \bibinfo {author} {\bibfnamefont {A.}~\bibnamefont
  {Gabrielli}},\ }\href {\doibase 10.1103/PhysRevLett.93.098501} {\bibfield
  {journal} {\bibinfo  {journal} {Physical Review Letters}\ }\textbf {\bibinfo
  {volume} {93}},\ \bibinfo {pages} {098501} (\bibinfo {year}
  {2004})}\BibitemShut {NoStop}%
\bibitem [{\citenamefont {Maritan}\ \emph {et~al.}(1996)\citenamefont
  {Maritan}, \citenamefont {Colaiori}, \citenamefont {Flammini}, \citenamefont
  {Cieplak},\ and\ \citenamefont {Banavar}}]{maritan_universality_1996}%
  \BibitemOpen
  \bibfield  {author} {\bibinfo {author} {\bibfnamefont {A.}~\bibnamefont
  {Maritan}}, \bibinfo {author} {\bibfnamefont {F.}~\bibnamefont {Colaiori}},
  \bibinfo {author} {\bibfnamefont {A.}~\bibnamefont {Flammini}}, \bibinfo
  {author} {\bibfnamefont {M.}~\bibnamefont {Cieplak}}, \ and\ \bibinfo
  {author} {\bibfnamefont {J.~R.}\ \bibnamefont {Banavar}},\ }\href {\doibase
  10.1126/science.272.5264.984} {\bibfield  {journal} {\bibinfo  {journal}
  {Science}\ }\textbf {\bibinfo {volume} {272}},\ \bibinfo {pages} {984}
  (\bibinfo {year} {1996})}\BibitemShut {NoStop}%
\bibitem [{\citenamefont {Gill}(2012)}]{gill2012orogenic}%
  \BibitemOpen
  \bibfield  {author} {\bibinfo {author} {\bibfnamefont {J.~B.}\ \bibnamefont
  {Gill}},\ }\href@noop {} {\emph {\bibinfo {title} {Orogenic andesites and
  plate tectonics}}}\ (\bibinfo  {publisher} {Springer Science \& Business
  Media},\ \bibinfo {year} {2012})\BibitemShut {NoStop}%
\bibitem [{\citenamefont {Albert}\ and\ \citenamefont
  {Barab{\'a}si}(2002)}]{albert_statistical_2002}%
  \BibitemOpen
  \bibfield  {author} {\bibinfo {author} {\bibfnamefont {R.}~\bibnamefont
  {Albert}}\ and\ \bibinfo {author} {\bibfnamefont {A.~L.}\ \bibnamefont
  {Barab{\'a}si}},\ }\href@noop {} {\bibfield  {journal} {\bibinfo  {journal}
  {Reviews of Modern Physics}\ }\textbf {\bibinfo {volume} {74}},\ \bibinfo
  {pages} {47} (\bibinfo {year} {2002})}\BibitemShut {NoStop}%
\bibitem [{\citenamefont {Cohen}\ and\ \citenamefont
  {Havlin}(2010)}]{cohen2010complex}%
  \BibitemOpen
  \bibfield  {author} {\bibinfo {author} {\bibfnamefont {R.}~\bibnamefont
  {Cohen}}\ and\ \bibinfo {author} {\bibfnamefont {S.}~\bibnamefont {Havlin}},\
  }\href@noop {} {\emph {\bibinfo {title} {Complex networks: structure,
  robustness and function}}}\ (\bibinfo  {publisher} {Cambridge university
  press},\ \bibinfo {year} {2010})\BibitemShut {NoStop}%
\bibitem [{\citenamefont {Brockmann}\ and\ \citenamefont
  {Helbing}(2013)}]{brockmann_hidden_2013}%
  \BibitemOpen
  \bibfield  {author} {\bibinfo {author} {\bibfnamefont {D.}~\bibnamefont
  {Brockmann}}\ and\ \bibinfo {author} {\bibfnamefont {D.}~\bibnamefont
  {Helbing}},\ }\href {\doibase 10.1126/science.1245200} {\bibfield  {journal}
  {\bibinfo  {journal} {Science}\ }\textbf {\bibinfo {volume} {342}},\ \bibinfo
  {pages} {1337} (\bibinfo {year} {2013})}\BibitemShut {NoStop}%
\bibitem [{\citenamefont {Newman}(2010)}]{newman2010networks}%
  \BibitemOpen
  \bibfield  {author} {\bibinfo {author} {\bibfnamefont {M.}~\bibnamefont
  {Newman}},\ }\href@noop {} {\emph {\bibinfo {title} {Networks: an
  introduction}}}\ (\bibinfo  {publisher} {Oxford university press},\ \bibinfo
  {year} {2010})\BibitemShut {NoStop}%
\bibitem [{\citenamefont {{Romualdo Pastor-Satorras}}\ \emph
  {et~al.}(2015)\citenamefont {{Romualdo Pastor-Satorras}}, \citenamefont
  {Castellano}, \citenamefont {Van~Mieghem},\ and\ \citenamefont
  {Vespignani}}]{romualdo_pastor-satorras_epidemic_2015}%
  \BibitemOpen
  \bibfield  {author} {\bibinfo {author} {\bibnamefont {{Romualdo
  Pastor-Satorras}}}, \bibinfo {author} {\bibfnamefont {C.}~\bibnamefont
  {Castellano}}, \bibinfo {author} {\bibfnamefont {P.}~\bibnamefont
  {Van~Mieghem}}, \ and\ \bibinfo {author} {\bibfnamefont {A.}~\bibnamefont
  {Vespignani}},\ }\href {\doibase 10.1103/RevModPhys.87.925} {\bibfield
  {journal} {\bibinfo  {journal} {Reviews of Modern Physics}\ }\textbf
  {\bibinfo {volume} {87}},\ \bibinfo {pages} {925} (\bibinfo {year}
  {2015})}\BibitemShut {NoStop}%
\bibitem [{\citenamefont {Tsonis}\ and\ \citenamefont
  {Swanson}(2008)}]{tsonis_topology_2008}%
  \BibitemOpen
  \bibfield  {author} {\bibinfo {author} {\bibfnamefont {A.~A.}\ \bibnamefont
  {Tsonis}}\ and\ \bibinfo {author} {\bibfnamefont {K.~L.}\ \bibnamefont
  {Swanson}},\ }\href {\doibase 10.1103/PhysRevLett.100.228502} {\bibfield
  {journal} {\bibinfo  {journal} {Physical Review Letters}\ }\textbf {\bibinfo
  {volume} {100}},\ \bibinfo {pages} {228502} (\bibinfo {year}
  {2008})}\BibitemShut {NoStop}%
\bibitem [{\citenamefont {Yamasaki}\ \emph {et~al.}(2008)\citenamefont
  {Yamasaki}, \citenamefont {Gozolchiani},\ and\ \citenamefont
  {Havlin}}]{yamasaki_climate_2008}%
  \BibitemOpen
  \bibfield  {author} {\bibinfo {author} {\bibfnamefont {K.}~\bibnamefont
  {Yamasaki}}, \bibinfo {author} {\bibfnamefont {A.}~\bibnamefont
  {Gozolchiani}}, \ and\ \bibinfo {author} {\bibfnamefont {S.}~\bibnamefont
  {Havlin}},\ }\href {\doibase 10.1103/PhysRevLett.100.228501} {\bibfield
  {journal} {\bibinfo  {journal} {Physical Review Letters}\ }\textbf {\bibinfo
  {volume} {100}},\ \bibinfo {pages} {228501} (\bibinfo {year}
  {2008})}\BibitemShut {NoStop}%
\bibitem [{\citenamefont {Donges}\ \emph
  {et~al.}(2009{\natexlab{a}})\citenamefont {Donges}, \citenamefont {Zou},
  \citenamefont {Marwan},\ and\ \citenamefont {Kurths}}]{donges_complex_2009}%
  \BibitemOpen
  \bibfield  {author} {\bibinfo {author} {\bibfnamefont {J.~F.}\ \bibnamefont
  {Donges}}, \bibinfo {author} {\bibfnamefont {Y.}~\bibnamefont {Zou}},
  \bibinfo {author} {\bibfnamefont {N.}~\bibnamefont {Marwan}}, \ and\ \bibinfo
  {author} {\bibfnamefont {J.}~\bibnamefont {Kurths}},\ }\href {\doibase
  10.1140/epjst/e2009-01098-2} {\bibfield  {journal} {\bibinfo  {journal} {The
  European Physical Journal Special Topics}\ }\textbf {\bibinfo {volume}
  {174}},\ \bibinfo {pages} {157} (\bibinfo {year}
  {2009}{\natexlab{a}})}\BibitemShut {NoStop}%
\bibitem [{\citenamefont {Donges}\ \emph
  {et~al.}(2009{\natexlab{b}})\citenamefont {Donges}, \citenamefont {Zou},
  \citenamefont {Marwan},\ and\ \citenamefont {Kurths}}]{donges_backbone_2009}%
  \BibitemOpen
  \bibfield  {author} {\bibinfo {author} {\bibfnamefont {J.~F.}\ \bibnamefont
  {Donges}}, \bibinfo {author} {\bibfnamefont {Y.}~\bibnamefont {Zou}},
  \bibinfo {author} {\bibfnamefont {N.}~\bibnamefont {Marwan}}, \ and\ \bibinfo
  {author} {\bibfnamefont {J.}~\bibnamefont {Kurths}},\ }\href {\doibase
  10.1209/0295-5075/87/48007} {\bibfield  {journal} {\bibinfo  {journal} {EPL
  (Europhysics Letters)}\ }\textbf {\bibinfo {volume} {87}},\ \bibinfo {pages}
  {48007} (\bibinfo {year} {2009}{\natexlab{b}})}\BibitemShut {NoStop}%
\bibitem [{\citenamefont {Barreiro}\ \emph {et~al.}(2011)\citenamefont
  {Barreiro}, \citenamefont {Marti},\ and\ \citenamefont
  {Masoller}}]{barreiro_inferring_2011}%
  \BibitemOpen
  \bibfield  {author} {\bibinfo {author} {\bibfnamefont {M.}~\bibnamefont
  {Barreiro}}, \bibinfo {author} {\bibfnamefont {A.~C.}\ \bibnamefont {Marti}},
  \ and\ \bibinfo {author} {\bibfnamefont {C.}~\bibnamefont {Masoller}},\
  }\href {\doibase 10.1063/1.3545273} {\bibfield  {journal} {\bibinfo
  {journal} {Chaos: An Interdisciplinary Journal of Nonlinear Science}\
  }\textbf {\bibinfo {volume} {21}},\ \bibinfo {pages} {013101} (\bibinfo
  {year} {2011})}\BibitemShut {NoStop}%
\bibitem [{\citenamefont {Martin}\ \emph {et~al.}(2013)\citenamefont {Martin},
  \citenamefont {Paczuski},\ and\ \citenamefont
  {Davidsen}}]{martin_interpretation_2013}%
  \BibitemOpen
  \bibfield  {author} {\bibinfo {author} {\bibfnamefont {E.~A.}\ \bibnamefont
  {Martin}}, \bibinfo {author} {\bibfnamefont {M.}~\bibnamefont {Paczuski}}, \
  and\ \bibinfo {author} {\bibfnamefont {J.}~\bibnamefont {Davidsen}},\ }\href
  {\doibase 10.1209/0295-5075/102/48003} {\bibfield  {journal} {\bibinfo
  {journal} {EPL (Europhysics Letters)}\ }\textbf {\bibinfo {volume} {102}},\
  \bibinfo {pages} {48003} (\bibinfo {year} {2013})}\BibitemShut {NoStop}%
\bibitem [{\citenamefont {Fan}\ \emph {et~al.}(2017)\citenamefont {Fan},
  \citenamefont {Meng}, \citenamefont {Ashkenazy}, \citenamefont {Havlin},\
  and\ \citenamefont {Schellnhuber}}]{fan2017network}%
  \BibitemOpen
  \bibfield  {author} {\bibinfo {author} {\bibfnamefont {J.}~\bibnamefont
  {Fan}}, \bibinfo {author} {\bibfnamefont {J.}~\bibnamefont {Meng}}, \bibinfo
  {author} {\bibfnamefont {Y.}~\bibnamefont {Ashkenazy}}, \bibinfo {author}
  {\bibfnamefont {S.}~\bibnamefont {Havlin}}, \ and\ \bibinfo {author}
  {\bibfnamefont {H.~J.}\ \bibnamefont {Schellnhuber}},\ }\href@noop {}
  {\bibfield  {journal} {\bibinfo  {journal} {Proceedings of the National
  Academy of Sciences}\ }\textbf {\bibinfo {volume} {114}},\ \bibinfo {pages}
  {7543–} (\bibinfo {year} {2017})}\BibitemShut {NoStop}%
\bibitem [{\citenamefont {Ludescher}\ \emph {et~al.}(2014)\citenamefont
  {Ludescher}, \citenamefont {Gozolchiani}, \citenamefont {Bogachev},
  \citenamefont {Bunde}, \citenamefont {Havlin},\ and\ \citenamefont
  {Schellnhuber}}]{ludescher_very_2014}%
  \BibitemOpen
  \bibfield  {author} {\bibinfo {author} {\bibfnamefont {J.}~\bibnamefont
  {Ludescher}}, \bibinfo {author} {\bibfnamefont {A.}~\bibnamefont
  {Gozolchiani}}, \bibinfo {author} {\bibfnamefont {M.~I.}\ \bibnamefont
  {Bogachev}}, \bibinfo {author} {\bibfnamefont {A.}~\bibnamefont {Bunde}},
  \bibinfo {author} {\bibfnamefont {S.}~\bibnamefont {Havlin}}, \ and\ \bibinfo
  {author} {\bibfnamefont {H.~J.}\ \bibnamefont {Schellnhuber}},\ }\href
  {\doibase 10.1073/pnas.1323058111} {\bibfield  {journal} {\bibinfo  {journal}
  {Proceedings of the National Academy of Sciences}\ }\textbf {\bibinfo
  {volume} {111}},\ \bibinfo {pages} {2064} (\bibinfo {year}
  {2014})}\BibitemShut {NoStop}%
\bibitem [{\citenamefont {Boers}\ \emph {et~al.}(2014)\citenamefont {Boers},
  \citenamefont {Bookhagen}, \citenamefont {Barbosa}, \citenamefont {Marwan},
  \citenamefont {Kurths},\ and\ \citenamefont
  {Marengo}}]{boers_prediction_2014}%
  \BibitemOpen
  \bibfield  {author} {\bibinfo {author} {\bibfnamefont {N.}~\bibnamefont
  {Boers}}, \bibinfo {author} {\bibfnamefont {B.}~\bibnamefont {Bookhagen}},
  \bibinfo {author} {\bibfnamefont {H.~M.~J.}\ \bibnamefont {Barbosa}},
  \bibinfo {author} {\bibfnamefont {N.}~\bibnamefont {Marwan}}, \bibinfo
  {author} {\bibfnamefont {J.}~\bibnamefont {Kurths}}, \ and\ \bibinfo {author}
  {\bibfnamefont {J.~A.}\ \bibnamefont {Marengo}},\ }\href {\doibase
  10.1038/ncomms6199} {\bibfield  {journal} {\bibinfo  {journal} {Nature
  Communications}\ }\textbf {\bibinfo {volume} {5}},\ \bibinfo {pages} {5199}
  (\bibinfo {year} {2014})}\BibitemShut {NoStop}%
\bibitem [{\citenamefont {Meng}\ \emph {et~al.}(2017)\citenamefont {Meng},
  \citenamefont {Fan}, \citenamefont {Ashkenazy},\ and\ \citenamefont
  {Havlin}}]{meng_percolation_2017}%
  \BibitemOpen
  \bibfield  {author} {\bibinfo {author} {\bibfnamefont {J.}~\bibnamefont
  {Meng}}, \bibinfo {author} {\bibfnamefont {J.}~\bibnamefont {Fan}}, \bibinfo
  {author} {\bibfnamefont {Y.}~\bibnamefont {Ashkenazy}}, \ and\ \bibinfo
  {author} {\bibfnamefont {S.}~\bibnamefont {Havlin}},\ }\href {\doibase
  10.1063/1.4975766} {\bibfield  {journal} {\bibinfo  {journal} {Chaos}\
  }\textbf {\bibinfo {volume} {27}},\ \bibinfo {pages} {035807} (\bibinfo
  {year} {2017})}\BibitemShut {NoStop}%
\bibitem [{\citenamefont {Meng}\ \emph {et~al.}(2018)\citenamefont {Meng},
  \citenamefont {Fan}, \citenamefont {Ashkenazy}, \citenamefont {Bunde},\ and\
  \citenamefont {Havlin}}]{meng_forecasting_2018}%
  \BibitemOpen
  \bibfield  {author} {\bibinfo {author} {\bibfnamefont {J.}~\bibnamefont
  {Meng}}, \bibinfo {author} {\bibfnamefont {J.}~\bibnamefont {Fan}}, \bibinfo
  {author} {\bibfnamefont {Y.}~\bibnamefont {Ashkenazy}}, \bibinfo {author}
  {\bibfnamefont {A.}~\bibnamefont {Bunde}}, \ and\ \bibinfo {author}
  {\bibfnamefont {S.}~\bibnamefont {Havlin}},\ }\href {\doibase
  10.1088/1367-2630/aabb25} {\bibfield  {journal} {\bibinfo  {journal} {New
  Journal of Physics}\ }\textbf {\bibinfo {volume} {20}},\ \bibinfo {pages}
  {043036} (\bibinfo {year} {2018})}\BibitemShut {NoStop}%
\bibitem [{\citenamefont {Runge}\ \emph {et~al.}(2015)\citenamefont {Runge},
  \citenamefont {Petoukhov}, \citenamefont {Donges}, \citenamefont {Hlinka},
  \citenamefont {Jajcay}, \citenamefont {Vejmelka}, \citenamefont {Hartman},
  \citenamefont {Marwan}, \citenamefont {Paluš},\ and\ \citenamefont
  {Kurths}}]{runge_identifying_2015}%
  \BibitemOpen
  \bibfield  {author} {\bibinfo {author} {\bibfnamefont {J.}~\bibnamefont
  {Runge}}, \bibinfo {author} {\bibfnamefont {V.}~\bibnamefont {Petoukhov}},
  \bibinfo {author} {\bibfnamefont {J.~F.}\ \bibnamefont {Donges}}, \bibinfo
  {author} {\bibfnamefont {J.}~\bibnamefont {Hlinka}}, \bibinfo {author}
  {\bibfnamefont {N.}~\bibnamefont {Jajcay}}, \bibinfo {author} {\bibfnamefont
  {M.}~\bibnamefont {Vejmelka}}, \bibinfo {author} {\bibfnamefont
  {D.}~\bibnamefont {Hartman}}, \bibinfo {author} {\bibfnamefont
  {N.}~\bibnamefont {Marwan}}, \bibinfo {author} {\bibfnamefont
  {M.}~\bibnamefont {Paluš}}, \ and\ \bibinfo {author} {\bibfnamefont
  {J.}~\bibnamefont {Kurths}},\ }\href {\doibase 10.1038/ncomms9502} {\bibfield
   {journal} {\bibinfo  {journal} {Nature Communications}\ }\textbf {\bibinfo
  {volume} {6}},\ \bibinfo {pages} {8502} (\bibinfo {year} {2015})}\BibitemShut
  {NoStop}%
\bibitem [{\citenamefont {Kitsak}\ \emph {et~al.}(2010)\citenamefont {Kitsak},
  \citenamefont {Gallos}, \citenamefont {Havlin}, \citenamefont {Liljeros},
  \citenamefont {Muchnik}, \citenamefont {Stanley},\ and\ \citenamefont
  {Makse}}]{kitsak_identification_2010}%
  \BibitemOpen
  \bibfield  {author} {\bibinfo {author} {\bibfnamefont {M.}~\bibnamefont
  {Kitsak}}, \bibinfo {author} {\bibfnamefont {L.~K.}\ \bibnamefont {Gallos}},
  \bibinfo {author} {\bibfnamefont {S.}~\bibnamefont {Havlin}}, \bibinfo
  {author} {\bibfnamefont {F.}~\bibnamefont {Liljeros}}, \bibinfo {author}
  {\bibfnamefont {L.}~\bibnamefont {Muchnik}}, \bibinfo {author} {\bibfnamefont
  {H.~E.}\ \bibnamefont {Stanley}}, \ and\ \bibinfo {author} {\bibfnamefont
  {H.~A.}\ \bibnamefont {Makse}},\ }\href {\doibase 10.1038/nphys1746}
  {\bibfield  {journal} {\bibinfo  {journal} {Nature Physics}\ }\textbf
  {\bibinfo {volume} {6}},\ \bibinfo {pages} {888} (\bibinfo {year}
  {2010})}\BibitemShut {NoStop}%
\bibitem [{\citenamefont {Liu}\ \emph {et~al.}(2011)\citenamefont {Liu},
  \citenamefont {Slotine},\ and\ \citenamefont
  {Barabási}}]{liu_controllability_2011}%
  \BibitemOpen
  \bibfield  {author} {\bibinfo {author} {\bibfnamefont {Y.-Y.}\ \bibnamefont
  {Liu}}, \bibinfo {author} {\bibfnamefont {J.-J.}\ \bibnamefont {Slotine}}, \
  and\ \bibinfo {author} {\bibfnamefont {A.-L.}\ \bibnamefont {Barabási}},\
  }\href {\doibase 10.1038/nature10011} {\bibfield  {journal} {\bibinfo
  {journal} {Nature}\ }\textbf {\bibinfo {volume} {473}},\ \bibinfo {pages}
  {167} (\bibinfo {year} {2011})}\BibitemShut {NoStop}%
\bibitem [{\citenamefont {Lü}\ \emph {et~al.}(2016)\citenamefont {Lü},
  \citenamefont {Chen}, \citenamefont {Ren}, \citenamefont {Zhang},
  \citenamefont {Zhang},\ and\ \citenamefont {Zhou}}]{lu_vital_2016}%
  \BibitemOpen
  \bibfield  {author} {\bibinfo {author} {\bibfnamefont {L.}~\bibnamefont
  {Lü}}, \bibinfo {author} {\bibfnamefont {D.}~\bibnamefont {Chen}}, \bibinfo
  {author} {\bibfnamefont {X.-L.}\ \bibnamefont {Ren}}, \bibinfo {author}
  {\bibfnamefont {Q.-M.}\ \bibnamefont {Zhang}}, \bibinfo {author}
  {\bibfnamefont {Y.-C.}\ \bibnamefont {Zhang}}, \ and\ \bibinfo {author}
  {\bibfnamefont {T.}~\bibnamefont {Zhou}},\ }\href {\doibase
  10.1016/j.physrep.2016.06.007} {\bibfield  {journal} {\bibinfo  {journal}
  {Physics Reports}\ }\textbf {\bibinfo {volume} {650}},\ \bibinfo {pages} {1}
  (\bibinfo {year} {2016})}\BibitemShut {NoStop}%
\bibitem [{\citenamefont {Isichenko}(1992)}]{isichenko_percolation_1992}%
  \BibitemOpen
  \bibfield  {author} {\bibinfo {author} {\bibfnamefont {M.~B.}\ \bibnamefont
  {Isichenko}},\ }\href {\doibase 10.1103/RevModPhys.64.961} {\bibfield
  {journal} {\bibinfo  {journal} {Reviews of Modern Physics}\ }\textbf
  {\bibinfo {volume} {64}},\ \bibinfo {pages} {961} (\bibinfo {year}
  {1992})}\BibitemShut {NoStop}%
\bibitem [{\citenamefont {Bunde}\ and\ \citenamefont
  {Havlin}(2012)}]{bunde2012fractals}%
  \BibitemOpen
  \bibfield  {author} {\bibinfo {author} {\bibfnamefont {A.}~\bibnamefont
  {Bunde}}\ and\ \bibinfo {author} {\bibfnamefont {S.}~\bibnamefont {Havlin}},\
  }\href@noop {} {\emph {\bibinfo {title} {Fractals and disordered systems}}}\
  (\bibinfo  {publisher} {Springer Science \& Business Media},\ \bibinfo {year}
  {2012})\BibitemShut {NoStop}%
\bibitem [{\citenamefont {Cohen}\ \emph {et~al.}(2000)\citenamefont {Cohen},
  \citenamefont {Erez}, \citenamefont {ben Avraham},\ and\ \citenamefont
  {Havlin}}]{cohen_resilience_2000}%
  \BibitemOpen
  \bibfield  {author} {\bibinfo {author} {\bibfnamefont {R.}~\bibnamefont
  {Cohen}}, \bibinfo {author} {\bibfnamefont {K.}~\bibnamefont {Erez}},
  \bibinfo {author} {\bibfnamefont {D.}~\bibnamefont {ben Avraham}}, \ and\
  \bibinfo {author} {\bibfnamefont {S.}~\bibnamefont {Havlin}},\ }\href
  {\doibase 10.1103/PhysRevLett.85.4626} {\bibfield  {journal} {\bibinfo
  {journal} {Physical Review Letters}\ }\textbf {\bibinfo {volume} {85}},\
  \bibinfo {pages} {4626} (\bibinfo {year} {2000})}\BibitemShut {NoStop}%
\bibitem [{\citenamefont {Stauffer}\ and\ \citenamefont
  {Aharony}(2003)}]{aharony2003introduction}%
  \BibitemOpen
  \bibfield  {author} {\bibinfo {author} {\bibfnamefont {D.}~\bibnamefont
  {Stauffer}}\ and\ \bibinfo {author} {\bibfnamefont {A.}~\bibnamefont
  {Aharony}},\ }\href@noop {} {\emph {\bibinfo {title} {Introduction to
  percolation theory}}}\ (\bibinfo  {publisher} {Taylor \& Francis},\ \bibinfo
  {year} {2003})\BibitemShut {NoStop}%
\bibitem [{\citenamefont {Taubert}\ \emph {et~al.}(2018)\citenamefont
  {Taubert}, \citenamefont {Fischer}, \citenamefont {Groeneveld}, \citenamefont
  {Lehmann}, \citenamefont {Müller}, \citenamefont {Rödig}, \citenamefont
  {Wiegand},\ and\ \citenamefont {Huth}}]{taubert_global_2018}%
  \BibitemOpen
  \bibfield  {author} {\bibinfo {author} {\bibfnamefont {F.}~\bibnamefont
  {Taubert}}, \bibinfo {author} {\bibfnamefont {R.}~\bibnamefont {Fischer}},
  \bibinfo {author} {\bibfnamefont {J.}~\bibnamefont {Groeneveld}}, \bibinfo
  {author} {\bibfnamefont {S.}~\bibnamefont {Lehmann}}, \bibinfo {author}
  {\bibfnamefont {M.~S.}\ \bibnamefont {Müller}}, \bibinfo {author}
  {\bibfnamefont {E.}~\bibnamefont {Rödig}}, \bibinfo {author} {\bibfnamefont
  {T.}~\bibnamefont {Wiegand}}, \ and\ \bibinfo {author} {\bibfnamefont
  {A.}~\bibnamefont {Huth}},\ }\href {\doibase 10.1038/nature25508} {\bibfield
  {journal} {\bibinfo  {journal} {Nature}\ } (\bibinfo {year} {2018}),\
  10.1038/nature25508}\BibitemShut {NoStop}%
\bibitem [{\citenamefont {Ali~Saberi}(2013)}]{ali_saberi_percolation_2013}%
  \BibitemOpen
  \bibfield  {author} {\bibinfo {author} {\bibfnamefont {A.}~\bibnamefont
  {Ali~Saberi}},\ }\href {\doibase 10.1103/PhysRevLett.110.178501} {\bibfield
  {journal} {\bibinfo  {journal} {Physical Review Letters}\ }\textbf {\bibinfo
  {volume} {110}},\ \bibinfo {pages} {178501} (\bibinfo {year}
  {2013})}\BibitemShut {NoStop}%
\bibitem [{\citenamefont {Li}\ \emph {et~al.}(2015)\citenamefont {Li},
  \citenamefont {Fu}, \citenamefont {Wang}, \citenamefont {Lu}, \citenamefont
  {Berezin}, \citenamefont {Stanley},\ and\ \citenamefont
  {Havlin}}]{li_percolation_2015}%
  \BibitemOpen
  \bibfield  {author} {\bibinfo {author} {\bibfnamefont {D.}~\bibnamefont
  {Li}}, \bibinfo {author} {\bibfnamefont {B.}~\bibnamefont {Fu}}, \bibinfo
  {author} {\bibfnamefont {Y.}~\bibnamefont {Wang}}, \bibinfo {author}
  {\bibfnamefont {G.}~\bibnamefont {Lu}}, \bibinfo {author} {\bibfnamefont
  {Y.}~\bibnamefont {Berezin}}, \bibinfo {author} {\bibfnamefont {H.~E.}\
  \bibnamefont {Stanley}}, \ and\ \bibinfo {author} {\bibfnamefont
  {S.}~\bibnamefont {Havlin}},\ }\href {\doibase 10.1073/pnas.1419185112}
  {\bibfield  {journal} {\bibinfo  {journal} {Proceedings of the National
  Academy of Sciences}\ }\textbf {\bibinfo {volume} {112}},\ \bibinfo {pages}
  {669} (\bibinfo {year} {2015})}\BibitemShut {NoStop}%
\bibitem [{\citenamefont {Morone}\ and\ \citenamefont
  {Makse}(2015)}]{morone_influence_2015}%
  \BibitemOpen
  \bibfield  {author} {\bibinfo {author} {\bibfnamefont {F.}~\bibnamefont
  {Morone}}\ and\ \bibinfo {author} {\bibfnamefont {H.~A.}\ \bibnamefont
  {Makse}},\ }\href {\doibase 10.1038/nature14604} {\bibfield  {journal}
  {\bibinfo  {journal} {Nature}\ }\textbf {\bibinfo {volume} {524}},\ \bibinfo
  {pages} {65} (\bibinfo {year} {2015})}\BibitemShut {NoStop}%
\bibitem [{\citenamefont {Saberi}(2015)}]{saberi2015recent}%
  \BibitemOpen
  \bibfield  {author} {\bibinfo {author} {\bibfnamefont {A.~A.}\ \bibnamefont
  {Saberi}},\ }\href@noop {} {\bibfield  {journal} {\bibinfo  {journal}
  {Physics Reports}\ }\textbf {\bibinfo {volume} {578}},\ \bibinfo {pages} {1}
  (\bibinfo {year} {2015})}\BibitemShut {NoStop}%
\bibitem [{\citenamefont {Amante}\ and\ \citenamefont
  {Eakins}()}]{amante1noaa}%
  \BibitemOpen
  \bibfield  {author} {\bibinfo {author} {\bibfnamefont {C.}~\bibnamefont
  {Amante}}\ and\ \bibinfo {author} {\bibfnamefont {B.}~\bibnamefont
  {Eakins}},\ }\href@noop {} {\enquote {\bibinfo {title} {Noaa technical
  memorandum nesdis ngdc-24; 2009},}\ }\BibitemShut {NoStop}%
\bibitem [{\citenamefont {Hoshen}\ and\ \citenamefont
  {Kopelman}(1976)}]{hoshen_percolation_1976}%
  \BibitemOpen
  \bibfield  {author} {\bibinfo {author} {\bibfnamefont {J.}~\bibnamefont
  {Hoshen}}\ and\ \bibinfo {author} {\bibfnamefont {R.}~\bibnamefont
  {Kopelman}},\ }\href {\doibase 10.1103/PhysRevB.14.3438} {\bibfield
  {journal} {\bibinfo  {journal} {Physical Review B}\ }\textbf {\bibinfo
  {volume} {14}},\ \bibinfo {pages} {3438} (\bibinfo {year}
  {1976})}\BibitemShut {NoStop}%
\bibitem [{\citenamefont {Newman}\ and\ \citenamefont
  {Ziff}(2000)}]{newman_efficient_2000}%
  \BibitemOpen
  \bibfield  {author} {\bibinfo {author} {\bibfnamefont {M.~E.~J.}\
  \bibnamefont {Newman}}\ and\ \bibinfo {author} {\bibfnamefont {R.~M.}\
  \bibnamefont {Ziff}},\ }\href {\doibase 10.1103/PhysRevLett.85.4104}
  {\bibfield  {journal} {\bibinfo  {journal} {Physical Review Letters}\
  }\textbf {\bibinfo {volume} {85}},\ \bibinfo {pages} {4104} (\bibinfo {year}
  {2000})}\BibitemShut {NoStop}%
\bibitem [{\citenamefont {Fan}\ \emph {et~al.}(2018)\citenamefont {Fan},
  \citenamefont {Meng}, \citenamefont {Ashkenazy}, \citenamefont {Havlin},\
  and\ \citenamefont {Schellnhuber}}]{fan2018climate}%
  \BibitemOpen
  \bibfield  {author} {\bibinfo {author} {\bibfnamefont {J.}~\bibnamefont
  {Fan}}, \bibinfo {author} {\bibfnamefont {J.}~\bibnamefont {Meng}}, \bibinfo
  {author} {\bibfnamefont {Y.}~\bibnamefont {Ashkenazy}}, \bibinfo {author}
  {\bibfnamefont {S.}~\bibnamefont {Havlin}}, \ and\ \bibinfo {author}
  {\bibfnamefont {H.~J.}\ \bibnamefont {Schellnhuber}},\ }\href {\doibase
  10.1073/pnas.1811068115} {\bibfield  {journal} {\bibinfo  {journal}
  {Proceedings of the National Academy of Sciences}\ }\textbf {\bibinfo
  {volume} {115}},\ \bibinfo {pages} {E12128} (\bibinfo {year}
  {2018})}\BibitemShut {NoStop}%
\bibitem [{\citenamefont {Wegener}(1966)}]{wegener1966origin}%
  \BibitemOpen
  \bibfield  {author} {\bibinfo {author} {\bibfnamefont {A.}~\bibnamefont
  {Wegener}},\ }\href@noop {} {\emph {\bibinfo {title} {The origin of
  continents and oceans}}}\ (\bibinfo  {publisher} {Courier Corporation},\
  \bibinfo {year} {1966})\BibitemShut {NoStop}%
\bibitem [{SI()}]{SI}%
  \BibitemOpen
  \href@noop {} {\bibinfo  {journal} {Supplementary materials}\ }\BibitemShut
  {NoStop}%
\bibitem [{\citenamefont {Bollob{\'a}s}(2001)}]{bollobas2001random}%
  \BibitemOpen
\bibfield  {journal} {  }\bibfield  {author} {\bibinfo {author} {\bibfnamefont
  {B.}~\bibnamefont {Bollob{\'a}s}},\ }\href@noop {} {\emph {\bibinfo {title}
  {Random graphs}}}\ (\bibinfo  {publisher} {Cambridge University Press},\
  \bibinfo {year} {2001})\BibitemShut {NoStop}%
\bibitem [{\citenamefont {Buldyrev}\ \emph {et~al.}(2010)\citenamefont
  {Buldyrev}, \citenamefont {Parshani}, \citenamefont {Paul}, \citenamefont
  {Stanley},\ and\ \citenamefont {Havlin}}]{buldyrev_catastrophic_2010}%
  \BibitemOpen
  \bibfield  {author} {\bibinfo {author} {\bibfnamefont {S.~V.}\ \bibnamefont
  {Buldyrev}}, \bibinfo {author} {\bibfnamefont {R.}~\bibnamefont {Parshani}},
  \bibinfo {author} {\bibfnamefont {G.}~\bibnamefont {Paul}}, \bibinfo {author}
  {\bibfnamefont {H.~E.}\ \bibnamefont {Stanley}}, \ and\ \bibinfo {author}
  {\bibfnamefont {S.}~\bibnamefont {Havlin}},\ }\href {\doibase
  10.1038/nature08932} {\bibfield  {journal} {\bibinfo  {journal} {Nature}\
  }\textbf {\bibinfo {volume} {464}},\ \bibinfo {pages} {1025} (\bibinfo {year}
  {2010})}\BibitemShut {NoStop}%
\bibitem [{\citenamefont {Hu}\ \emph {et~al.}(2011)\citenamefont {Hu},
  \citenamefont {Ksherim}, \citenamefont {Cohen},\ and\ \citenamefont
  {Havlin}}]{hu_percolation_2011}%
  \BibitemOpen
  \bibfield  {author} {\bibinfo {author} {\bibfnamefont {Y.}~\bibnamefont
  {Hu}}, \bibinfo {author} {\bibfnamefont {B.}~\bibnamefont {Ksherim}},
  \bibinfo {author} {\bibfnamefont {R.}~\bibnamefont {Cohen}}, \ and\ \bibinfo
  {author} {\bibfnamefont {S.}~\bibnamefont {Havlin}},\ }\href {\doibase
  10.1103/PhysRevE.84.066116} {\bibfield  {journal} {\bibinfo  {journal}
  {Physical Review E}\ }\textbf {\bibinfo {volume} {84}},\ \bibinfo {pages}
  {066116} (\bibinfo {year} {2011})}\BibitemShut {NoStop}%
\bibitem [{\citenamefont {Gao}\ \emph {et~al.}(2012)\citenamefont {Gao},
  \citenamefont {Buldyrev}, \citenamefont {Stanley},\ and\ \citenamefont
  {Havlin}}]{gao_networks_2012}%
  \BibitemOpen
  \bibfield  {author} {\bibinfo {author} {\bibfnamefont {J.}~\bibnamefont
  {Gao}}, \bibinfo {author} {\bibfnamefont {S.~V.}\ \bibnamefont {Buldyrev}},
  \bibinfo {author} {\bibfnamefont {H.~E.}\ \bibnamefont {Stanley}}, \ and\
  \bibinfo {author} {\bibfnamefont {S.}~\bibnamefont {Havlin}},\ }\href
  {\doibase 10.1038/nphys2180} {\bibfield  {journal} {\bibinfo  {journal}
  {Nature Physics}\ }\textbf {\bibinfo {volume} {8}},\ \bibinfo {pages} {40}
  (\bibinfo {year} {2012})}\BibitemShut {NoStop}%
\bibitem [{\citenamefont {Achlioptas}\ \emph {et~al.}(2009)\citenamefont
  {Achlioptas}, \citenamefont {D'Souza},\ and\ \citenamefont
  {Spencer}}]{achlioptas_explosive_2009}%
  \BibitemOpen
  \bibfield  {author} {\bibinfo {author} {\bibfnamefont {D.}~\bibnamefont
  {Achlioptas}}, \bibinfo {author} {\bibfnamefont {R.~M.}\ \bibnamefont
  {D'Souza}}, \ and\ \bibinfo {author} {\bibfnamefont {J.}~\bibnamefont
  {Spencer}},\ }\href {\doibase 10.1126/science.1167782} {\bibfield  {journal}
  {\bibinfo  {journal} {Science}\ }\textbf {\bibinfo {volume} {323}},\ \bibinfo
  {pages} {1453} (\bibinfo {year} {2009})}\BibitemShut {NoStop}%
\bibitem [{\citenamefont {Riordan}\ and\ \citenamefont
  {Warnke}(2011)}]{riordan_explosive_2011}%
  \BibitemOpen
  \bibfield  {author} {\bibinfo {author} {\bibfnamefont {O.}~\bibnamefont
  {Riordan}}\ and\ \bibinfo {author} {\bibfnamefont {L.}~\bibnamefont
  {Warnke}},\ }\href {\doibase 10.1126/science.1206241} {\bibfield  {journal}
  {\bibinfo  {journal} {Science}\ }\textbf {\bibinfo {volume} {333}},\ \bibinfo
  {pages} {322} (\bibinfo {year} {2011})}\BibitemShut {NoStop}%
\bibitem [{\citenamefont {Fan}\ \emph {et~al.}(2012)\citenamefont {Fan},
  \citenamefont {Liu}, \citenamefont {Li},\ and\ \citenamefont
  {Chen}}]{fan_continuous_2012}%
  \BibitemOpen
  \bibfield  {author} {\bibinfo {author} {\bibfnamefont {J.}~\bibnamefont
  {Fan}}, \bibinfo {author} {\bibfnamefont {M.}~\bibnamefont {Liu}}, \bibinfo
  {author} {\bibfnamefont {L.}~\bibnamefont {Li}}, \ and\ \bibinfo {author}
  {\bibfnamefont {X.}~\bibnamefont {Chen}},\ }\href {\doibase
  10.1103/PhysRevE.85.061110} {\bibfield  {journal} {\bibinfo  {journal}
  {Physical Review E}\ }\textbf {\bibinfo {volume} {85}},\ \bibinfo {pages}
  {061110} (\bibinfo {year} {2012})}\BibitemShut {NoStop}%
\bibitem [{\citenamefont {D'Souza}\ and\ \citenamefont
  {Nagler}(2015)}]{dsouza_anomalous_2015}%
  \BibitemOpen
  \bibfield  {author} {\bibinfo {author} {\bibfnamefont {R.~M.}\ \bibnamefont
  {D'Souza}}\ and\ \bibinfo {author} {\bibfnamefont {J.}~\bibnamefont
  {Nagler}},\ }\href {\doibase 10.1038/nphys3378} {\bibfield  {journal}
  {\bibinfo  {journal} {Nature Physics}\ }\textbf {\bibinfo {volume} {11}},\
  \bibinfo {pages} {531} (\bibinfo {year} {2015})}\BibitemShut {NoStop}%
\bibitem [{\citenamefont {Nagler}\ \emph {et~al.}(2011)\citenamefont {Nagler},
  \citenamefont {Levina},\ and\ \citenamefont {Timme}}]{nagler_impact_2011}%
  \BibitemOpen
  \bibfield  {author} {\bibinfo {author} {\bibfnamefont {J.}~\bibnamefont
  {Nagler}}, \bibinfo {author} {\bibfnamefont {A.}~\bibnamefont {Levina}}, \
  and\ \bibinfo {author} {\bibfnamefont {M.}~\bibnamefont {Timme}},\ }\href
  {\doibase 10.1038/nphys1860} {\bibfield  {journal} {\bibinfo  {journal}
  {Nature Physics}\ }\textbf {\bibinfo {volume} {7}},\ \bibinfo {pages} {265}
  (\bibinfo {year} {2011})}\BibitemShut {NoStop}%
\bibitem [{\citenamefont {Mandelbrot}(1967)}]{mandelbrot_how_1967}%
  \BibitemOpen
  \bibfield  {author} {\bibinfo {author} {\bibfnamefont {B.}~\bibnamefont
  {Mandelbrot}},\ }\href {\doibase 10.1126/science.156.3775.636} {\bibfield
  {journal} {\bibinfo  {journal} {Science}\ }\textbf {\bibinfo {volume}
  {156}},\ \bibinfo {pages} {636} (\bibinfo {year} {1967})}\BibitemShut
  {NoStop}%
\bibitem [{\citenamefont {Du}\ \emph {et~al.}(1996)\citenamefont {Du},
  \citenamefont {Satik},\ and\ \citenamefont {Yortsos}}]{du_percolation_1996}%
  \BibitemOpen
  \bibfield  {author} {\bibinfo {author} {\bibfnamefont {C.}~\bibnamefont
  {Du}}, \bibinfo {author} {\bibfnamefont {C.}~\bibnamefont {Satik}}, \ and\
  \bibinfo {author} {\bibfnamefont {Y.~C.}\ \bibnamefont {Yortsos}},\ }\href
  {\doibase 10.1002/aic.690420831} {\bibfield  {journal} {\bibinfo  {journal}
  {AIChE Journal}\ }\textbf {\bibinfo {volume} {42}},\ \bibinfo {pages} {2392}
  (\bibinfo {year} {1996})}\BibitemShut {NoStop}%
\bibitem [{\citenamefont {Vicsek}(1992)}]{s1992fractal}%
  \BibitemOpen
  \bibfield  {author} {\bibinfo {author} {\bibfnamefont {T.}~\bibnamefont
  {Vicsek}},\ }\href@noop {} {\emph {\bibinfo {title} {Fractal growth
  phenomena}}}\ (\bibinfo  {publisher} {World scientific},\ \bibinfo {year}
  {1992})\BibitemShut {NoStop}%
\bibitem [{\citenamefont {Family}\ and\ \citenamefont
  {Vicsek}(1991)}]{family1991dynamics}%
  \BibitemOpen
  \bibfield  {author} {\bibinfo {author} {\bibfnamefont {F.}~\bibnamefont
  {Family}}\ and\ \bibinfo {author} {\bibfnamefont {T.}~\bibnamefont
  {Vicsek}},\ }\href@noop {} {\emph {\bibinfo {title} {Dynamics of fractal
  surfaces}}}\ (\bibinfo  {publisher} {World Scientific},\ \bibinfo {year}
  {1991})\BibitemShut {NoStop}%
\bibitem [{\citenamefont {Privman}(1990)}]{privman1990finite}%
  \BibitemOpen
  \bibfield  {author} {\bibinfo {author} {\bibfnamefont {V.}~\bibnamefont
  {Privman}},\ }\href@noop {} {\emph {\bibinfo {title} {Finite size scaling and
  numerical simulation of statistical systems}}}\ (\bibinfo  {publisher} {World
  Scientific Singapore},\ \bibinfo {year} {1990})\BibitemShut {NoStop}%
\bibitem [{\citenamefont {Castro}\ \emph {et~al.}(2018)\citenamefont {Castro},
  \citenamefont {Luković}, \citenamefont {Pompanin}, \citenamefont {Andrade},\
  and\ \citenamefont {Herrmann}}]{castro_schramm-loewner_2018}%
  \BibitemOpen
  \bibfield  {author} {\bibinfo {author} {\bibfnamefont {C.~P.~d.}\
  \bibnamefont {Castro}}, \bibinfo {author} {\bibfnamefont {M.}~\bibnamefont
  {Luković}}, \bibinfo {author} {\bibfnamefont {G.}~\bibnamefont {Pompanin}},
  \bibinfo {author} {\bibfnamefont {R.~F.~S.}\ \bibnamefont {Andrade}}, \ and\
  \bibinfo {author} {\bibfnamefont {H.~J.}\ \bibnamefont {Herrmann}},\ }\href
  {\doibase 10.1038/s41598-018-23489-x} {\bibfield  {journal} {\bibinfo
  {journal} {Scientific Reports}\ }\textbf {\bibinfo {volume} {8}},\ \bibinfo
  {pages} {5286} (\bibinfo {year} {2018})}\BibitemShut {NoStop}%
\bibitem [{\citenamefont {Schrenk}\ \emph {et~al.}(2013)\citenamefont
  {Schrenk}, \citenamefont {Posé}, \citenamefont {Kranz}, \citenamefont {van
  Kessenich}, \citenamefont {Araújo},\ and\ \citenamefont
  {Herrmann}}]{schrenk_percolation_2013}%
  \BibitemOpen
  \bibfield  {author} {\bibinfo {author} {\bibfnamefont {K.~J.}\ \bibnamefont
  {Schrenk}}, \bibinfo {author} {\bibfnamefont {N.}~\bibnamefont {Posé}},
  \bibinfo {author} {\bibfnamefont {J.~J.}\ \bibnamefont {Kranz}}, \bibinfo
  {author} {\bibfnamefont {L.~V.~M.}\ \bibnamefont {van Kessenich}}, \bibinfo
  {author} {\bibfnamefont {N.~A.~M.}\ \bibnamefont {Araújo}}, \ and\ \bibinfo
  {author} {\bibfnamefont {H.~J.}\ \bibnamefont {Herrmann}},\ }\href {\doibase
  10.1103/PhysRevE.88.052102} {\bibfield  {journal} {\bibinfo  {journal}
  {Physical Review E}\ }\textbf {\bibinfo {volume} {88}},\ \bibinfo {pages}
  {052102} (\bibinfo {year} {2013})}\BibitemShut {NoStop}%
\bibitem [{\citenamefont {Cathcart}\ \emph {et~al.}(2010)\citenamefont
  {Cathcart}, \citenamefont {Bolonkin},\ and\ \citenamefont
  {Rugescu}}]{cathcart2010bering}%
  \BibitemOpen
  \bibfield  {author} {\bibinfo {author} {\bibfnamefont {R.~B.}\ \bibnamefont
  {Cathcart}}, \bibinfo {author} {\bibfnamefont {A.~A.}\ \bibnamefont
  {Bolonkin}}, \ and\ \bibinfo {author} {\bibfnamefont {R.~D.}\ \bibnamefont
  {Rugescu}},\ }in\ \href@noop {} {\emph {\bibinfo {booktitle}
  {Macro-engineering Seawater in Unique Environments}}}\ (\bibinfo  {publisher}
  {Springer},\ \bibinfo {year} {2010})\ pp.\ \bibinfo {pages}
  {741--777}\BibitemShut {NoStop}%
\end{thebibliography}%

\end{document}